\documentclass[12pt]{article}
\usepackage{amssymb,amsmath,epsfig}
\usepackage{graphicx}
\allowdisplaybreaks

\begin{document}
\title{\bf Anisotropic Durgapal-Fuloria Neutron Stars in $f(\mathcal{R},\mathrm{T}^{2})$ Gravity}
\author{Tayyab Naseer \thanks{tayyabnaseer48@yahoo.com;
tayyab.naseer@math.uol.edu.pk}~, M. Sharif
\thanks{msharif.math@pu.edu.pk}~, Sana Manzoor
\thanks{sanamanzoormath@gmail.com}~ and Arooj Fatima \thanks{arooj3740@gmail.com}\\
Department of Mathematics and Statistics, The University of Lahore,\\
1-KM Defence Road Lahore-54000, Pakistan.}
\date{}
\maketitle

\begin{abstract}
The main purpose of this paper is to obtain physically stable
stellar models coupled with anisotropic matter distribution in the
context of $f(\mathcal{R},\mathrm{T}^{2})$ theory. For this, we
consider a static spherical geometry and formulate modified field
equations containing various unknowns such as matter determinants
and metric potentials. We then obtain a unique solution to these
equations by employing Durgapal-Fuloria ansatz possessing a constant
doublet. We also use matching criteria to calculate the values of
these constants by considering the Schwarzschild exterior spacetime.
Two different viable models of this modified theory are adopted to
analyze the behavior of effective matter variables, anisotropy,
energy conditions, compactness and redshift in the interiors of Her
X-1, PSR J0348-0432, LMC X-4, SMC X-1, Cen X-3, and SAX J
1808.4-3658 star candidates. We also check the stability of these
models by using three different physical tests. It is concluded that
our considered stars satisfy all the physical requirements and are
stable in this modified gravity for the considered parametric
values.
\end{abstract}

\section{Introduction}

General Relativity (GR) describes the fundamental forces that govern
the motion of objects in the universe, particularly, in the presence
of massive structures like stars and planets. This is an essential
part of the foundation of modern cosmology and our understanding of
cosmic structure as well as evolution. Singularities are the key
issues in GR because they represent points in spacetime where the
curvature becomes infinite and the laws of physics break down. In
specific scenarios, singularities are predicted to appear such as at
the core of a black hole or during the initial moments of the big
bang when energy densities and curvatures reach exceedingly high
levels. Consequently, the predictions made by GR regarding extreme
energy levels, particularly, the singularity linked to the origin of
the universe (the big bang) lose their validity. A reliable
explanation of these unusual instances has been provided by modified
theories of GR. Different approaches, i.e., higher-dimensional
gravity \cite{a}, $f(\mathcal{R})$ \cite{b,bb},
$f(\mathcal{R},\mathrm{T})$ \cite{c}-\cite{cb} and scalar-tensor
theories \cite{e} have been proposed and examined as a result of
this pursuit. These alternative theories enhance our understanding
of how gravity operates in scenarios where classical GR proves
insufficient.

In $2014$, Katirci and Kavuk \cite{f} presented a significant
modification to GR by including self-contraction of the
energy-momentum tensor, i.e.,
$(\mathrm{T}_{\xi\psi}\mathrm{T}^{\xi\psi}=\mathrm{T}^{2})$ into the
action. They named this modified theory as ``energy-momentum squared
gravity'' (EMSG) or $f(\mathcal{R},\mathrm{T}^{2})$ theory. This
theory potentially resolved significant cosmological puzzles as it
has been suggested as a candidate for explaining the observed
accelerated expansion of the universe. It also offers a promising
approach to understand the early-time universe in a more
comprehensive manner. An important feature of this theory is its
ability to calculate the additional acceleration that affects the
perihelion of Mercury. This is accomplished by determining the
Newtonian limit of the model. The field equations of this theory
involve squared and product terms of the matter variables, helping
in studying diverse cosmological scenarios. Another noteworthy
feature of this theory is that the traditional conservation of
energy and momentum may not hold due to the interaction between
matter and curvature that introduces an additional force. As a
result, the trajectory of a test particle differs from the standard
geodesic path predicted by GR.

This modified theory has sparked significant interest in the fields
of astrophysics and cosmology. Researchers have explored its
implications for various astronomical structures, seeking to
understand how this novel framework alters our understanding of the
universe evolution and behavior, particularly, in extreme conditions
where classical physics breaks down. The homogeneous and isotropic
spacetime has been studied in the modified
$f(\mathcal{R},\mathrm{T}^{2})$ theory and observed that there was a
possibility of bounce at early-times instead of the big bang
singularity. Board and Barrow \cite{i} found an exact solution for
isotropic universe and discussed their behavior with respect to the
early and late-time cosmic evolution in this theory. Moraes and
Sahoo \cite{j} proposed the existence of non-exotic matter wormholes
in this framework. The effect of this modified theory of gravity on
neutron stars and nuclear properties of the matter distribution has
been extensively discussed \cite{l}.

Celestial objects, particularly, stars are assumed as a fundamental
element in shaping the composition of galaxies within our universe.
The intricate structure of these celestial bodies has captivated the
attention of numerous astrophysicists, who have dedicated their
efforts to analyze the various stages of their evolution. The
gravitational collapse marks the end of a star, resulting in new
astrophysical structures such as black holes, neutron stars and
white dwarfs depending on the mass of the dying star. Neutron stars,
in particular, have garnered significant attention due to their
fascinating structural characteristics and evolutionary stages.
Numerous researchers have explored these structures and their
formation. Dev and Gleiser \cite{17} analyzed the surface redshift
and stability of self-gravitating stars. Mak and Harko \cite{18}
obtained a class of exact solutions, describing static spherically
symmetric stellar configuration. Kalam et al. \cite{19} used the
Krori-Barua metric to study the physical characteristics of strange
stars. A large body of literature exists on the discussion of
different properties of neutron stars \cite{20,20a}. Singh and Pant
\cite{21} found a class of exact solutions for anisotropic stars and
examined that all the energy conditions are satisfied for these
solutions. Maurya and Tello-Ortiz \cite{23} examined the anisotropy
and surface redshift of celestial bodies in the modified
$f(\mathcal{R})$ theory. Sharif and Naseer \cite{24}-\cite{24b}
employed a particular EoS to examine the stability of various
neutron star candidates in a non-minimal gravity. Some other
approaches have also been used in the literature to discuss the
neutron star models \cite{8a}-\cite{8h}.

A family of isotropic solutions, whether exact or approximate, does
not have physical relevance and is not aligned with what we observe
in astrophysical phenomena \cite{2}-\cite{4}. Researchers have found
compelling evidences that various intriguing physical phenomena can
lead to local deviations from uniformity and isotropy. These
deviations can occur within both low and high density regimes. For
example, in the later scenario, such as the cores of extremely
compact astrophysical objects where matter is packed incredibly
dense (densities even higher than that of the nuclear density
$3\times10^{17}kg/m^{3}$), exhibits anisotropic behavior \cite{5}.
Anisotropic behavior in these high-density objects occurs because
the pressure inside them does not act uniformly in all directions.
Instead, it can be divided into two distinct components, i.e.,
radial and transverse. Anisotropy in fluid pressure often emerges
from various factors and physical conditions within a system such as
the mixture of fluids of different types, rotation, viscosity, the
existence of a solid core, the presence of a superfluid or a
magnetic field \cite{18}. Researchers have extensively explored
these sources of anisotropy in higher dimensions \cite{10}-\cite{8}.
A body of literature is present that discusses anisotropic neutron
stars using different approaches \cite{62,63} as well as attempts to
understand the anisotropic pressure in such stellar objects from
matter/geometic aspects \cite{64}-\cite{67}. Some other recent
neutron stars observation reports from different collaborators can
be seen in \cite{68}-\cite{71}. Hence, it becomes appealing to apply
relativistic principles to understand how the fundamental forces and
particles interact, and affect the behavior of astrophysical
objects, contributing to broader our understanding of the cosmos.

Durgapal and Fuloria \cite{12} presented a viable perfect fluid
solution to characterize the incredibly dense configurations of
stellar objects like neutron stars. Their work provided a
theoretical framework that not only accurately represents the
extreme conditions within neutron stars but also ensures the
stability of this solution against radial perturbations. Later, this
solution was extended for anisotropic as well as charged stellar
configurations. Maurya et al. \cite{13} used this ansatz to obtain
viable star models coupled with the anisotropic matter distribution.
Komathiraj et al. \cite{14} generalized the isotropic
Durgapal-Fuloria solution for anisotropic charged stellar objects
and analyzed their physical properties. Maurya et al. \cite{16}
formulated the complexity-free anisotropic generalization of such an
isotropic model through the extended geometric deformations.

In this paper, we analyze anisotropic neutron stars with static
spherically symmetric interior in $f(\mathcal{R},\mathrm{T}^{2})$
theory. In section \textbf{2}, we take Durgapal-Fuloria ansatz and
construct the corresponding field equations. In order to calculate
the values of unknown constants, we use matching conditions between
the interior and the Schwarzschild exterior geometries. In sections
\textbf{3} and \textbf{4}, we choose different viable models of this
modified theory and explore physical properties of the considered
stars. We examine the effective matter variables, energy bounds, EoS
parameters corresponding to the resulting solutions. We also check
stability through the sound speed as well as the cracking approach
in section \textbf{5}. Section \textbf{6} summarizes our obtained
results.

\section{$f(\mathcal{R},\mathrm{T}^{2})$ Theory}

The Einstein-Hilbert action takes the form in EMSG scenario (for
$\kappa=1$) as \cite{l}
\begin{equation}\label{1}
I=\int
\sqrt{-g}\bigg\{\frac{f(\mathcal{R},\mathrm{T}^{2})}{2}+L_{m}\bigg\}
d^{4}x,
\end{equation}
where $L_{m}$ defines the matter Lagrangian density of the fluid
configuration and $g=|g_{\xi\psi}|$ with $g_{\xi\psi}$ being the
metric tensor. The field equations corresponding to the above
modified action are formulated as
\begin{equation}\label{2}
\mathcal{R}_{\xi\psi}f_{\mathcal{R}}+g _{\xi\psi}\nabla_{\xi}
\nabla^{\xi}f_{\mathcal{R}}-\nabla_{\xi}
\nabla_{\psi}f_{\mathcal{R}}-\frac{1}{2}g_{\xi\psi}f=\mathrm{T}_{\xi\psi}-\Theta_{\xi
\psi}f_{\mathrm{T}^{2}},
\end{equation}
where $f_{\mathrm{T}^{2}}=\frac{\partial
f(\mathcal{R},\mathrm{T}^{2})}{\partial \mathrm{T}^{2}}$ and
$f_{\mathcal{R}}=\frac{\partial
f(\mathcal{R},\mathrm{T}^{2})}{\partial\mathcal{R}}$. Also,
$\mathrm{T}_{\xi\psi}$ is the usual energy-momentum tensor and
\begin{equation}\label{3}
\Theta_{\xi\psi}=2\mathrm{T}^{\zeta}_{\xi}\mathrm{T}_{\psi\zeta}-2L_{m}\left(\mathrm{T}_{\xi
\psi}-\frac{1}{2}g_{\xi\psi}\mathrm{T}\right)-\frac{4\partial^{2}L_{m}}{\partial
g^{\xi\psi}\partial
g^{\zeta\eta}}\mathrm{T}^{\zeta\eta}-\mathrm{T}\mathrm{T}_{\xi\psi}.
\end{equation}
Rearranging Eq.(\ref{2}), we obtain
\begin{equation}\label{4}
G_{\xi\psi}=\mathcal{R}_{\xi\psi}-\frac{1}{2}\mathcal{R}g_{\xi
\psi}=\mathrm{T}^{eff}_{\xi\psi},
\end{equation}
where the EMSG corrections are given by
\begin{equation}\label{5}
\mathrm{T}^{eff}_{\xi\psi}=\frac{1}{f_{\mathcal{R}}}\big[\mathrm{T}_{\xi
\psi}+\nabla_{\xi}\nabla_{\psi}f_{\mathcal{R}}-g_{\xi
\psi}\nabla_{\xi}\nabla^{\xi}f_{\mathcal{R}}+\frac{1}{2}g_{\xi
\psi}(f-\mathcal{R}f_{\mathcal{R}})-\Theta_{\xi
\psi}f_{\mathrm{T}^{2}}\big].
\end{equation}

We consider that the hypersurface distinguishes the interior and
exterior region of a self-gravitating geometry. In order to study
the physical properties of stellar structures, we assume a static
sphere as the interior metric
\begin{equation}\label{6}
ds^{2}_{-}=-e^{\tau}dt^{2}+e^{\nu}dr^{2}+r^{2}\big(d\theta^{2}+\sin^{2}\theta
d\phi^{2}\big),
\end{equation}
where $\tau=\tau(r)$ and $\nu=\nu(r)$. The energy-momentum tensor
plays a crucial role in the modeling of celestial objects because it
specifies the interior matter distribution. In the realm of
astrophysics, the study of compact neutron-like stars has been a
fascinating area of exploration, providing valuable insights into
the fundamental nature of matter under extreme conditions.
Traditional approaches have predominantly focused on isotropic
models to understand these dense celestial objects. However, the
limitations of isotropic models become evident when faced with the
intricacies of celestial systems. This motivation seeks to highlight
the imperative shift towards investigating anisotropic stars within
the framework of modified theories, presenting a unique and
promising avenue for advancing our understanding of the cosmos.
Since we aim to model anisotropic star candidates, the corresponding
energy-momentum tensor has the form
\begin{equation}\label{7}
\mathrm{T}_{\xi\psi}=\rho U_{\xi}U_{\psi}+P_{r}
V_{\xi}V_{\psi}+P_{t}(U_{\xi}U_{\psi}-V_{\xi}V_{\psi}+g_{\xi\psi}),
\end{equation}
where the triplet $(P_{t},P_{r},\rho)$ symbolizes the
tangential/radial pressures and the energy density, respectively.
Also, $U_{\xi}$ is the four-velocity and $V_{\xi}$ indicates the
four-vector. It is mentioned here that a large body of literature
formulates physically relevant models for the Lagrangian
$L_{m}=\frac{P_{r}+2P_{t}}{3}$, thus we consider it in this case.
Joining this with the field equations \eqref{4}, we get
\begin{eqnarray}\nonumber
\rho^{eff}&=&\frac{1}{f_{\mathcal{R}}}\bigg[\rho+\frac{\mathcal{R}
f_{\mathcal{R}}-f}{2}+f_{\mathrm{T}^{2}}
\bigg\{\left(\frac{P_{r}+2P_{t}}{3}\right)(\rho+2P_{t}+P_{r})\\\label{8}
&+&\rho^{2}+2\rho P_{t}+\rho P_{r}\bigg\}+\frac{1}{e^{\nu}}
\bigg\{f''_{\mathcal{R}}-\bigg(\frac{\nu'}{2}
-\frac{2}{r}\bigg)f'_{\mathcal{R}}\bigg\}\bigg],
\\\nonumber
P^{eff}_{r}&=&\frac{1}{f_{\mathcal{R}}}\bigg[P_{r}-\frac{\mathcal{R}
f_{\mathcal{R}}-f}{2}+\bigg\{\left(\frac{P_{r}+2P_{t}}{3}\right)(P_{r}-2P_{t}+\rho)
\\\label{9}&-&(P^{2}_{r}-2P_{r}P_{t}+\rho P_{r})\bigg\}f_{\mathrm{T}^{2}}-\frac{1}{e^{\nu}}
\bigg(\frac{2}{r}+\frac{\tau'}{2}\bigg) f'_{\mathcal{R}}\bigg],
\\\nonumber
P^{eff}_{t}&=&\frac{1}{f_{\mathcal{R}}}\bigg[P_{t}-\frac{\mathcal{R}
f_{\mathcal{R}}-f}{2}-\frac{1}{e^{\nu}}\bigg\{f'_{\mathcal{R}}\bigg(\frac{\tau'}{2}
+\frac{1}{r}-\frac{\nu'}{2}\bigg)+f''_{\mathcal{R}}
\bigg\}\\\label{10}
&+&\bigg\{\left(\frac{P_{r}+2P_{t}}{3}\right)(\rho-P_{r})+P_{r}P_{t}-\rho
P_{t}\bigg\}f_{\mathrm{T}^{2}}\bigg].
\end{eqnarray}

We observe that there are five unknowns
($\nu,\tau,\rho,P_{r},P_{t}$) in the above three equations, making
them difficult to solve. In order to find a unique solution to these
equations, we adopt a particular form of the metric components in
the following. The singularity-free Durgapal-Fuloria ansatz is given
by \cite{12}
\begin{table}
\scriptsize \centering \caption{Estimated data of various stellar
objects and their corresponding calculated values of constants.}
\label{Table1} \vspace{+0.1in} \setlength{\tabcolsep}{1.4em}
\begin{tabular}{cccccc}
\hline\hline Star models & $M(M_{\odot})$ & $\mathrm{R}$ &
$M/\mathrm{R}$ & $A$ & $B$
\\\hline Her X-1 \cite{26} & 0.85 & 8.1 & 0.1049 & 0.606611 & 0.00104111
\\\hline
PSR J0348-0432 \cite{26} & 2.1 & 10.06 & 0.20874 & 0.3281 &
0.00152477
\\\hline
LMC X-4 \cite{28} & 1.29 & 8.831 & 0.14607 & 0.483925 & 0.00127895
\\\hline
SMC X-1 \cite{28} & 1.04 & 9.34 & 0.1113 &  0.58638 & 0.000836898
\\\hline
Cen X-3 \cite{28} & 1.49 & 10.8 & 0.1380 & 0.506798 & 0.000799845
\\\hline
SAX J 1808.4-3658 \cite{29} & 1.44 & 7.07 & 0.2036 &
0.339377 & 0.00299192\\
\hline\hline
\end{tabular}
\end{table}
\begin{eqnarray}\label{11}
e^{\tau(r)}&=&A(1+Br^{2})^{4}, \\\label{11a}
e^{\nu(r)}&=&\frac{7(1+Br^{2})^{2}}{7-B^{2}r^{4}-10Br^{2}},
\end{eqnarray}
where $A$ and $B$ are constants. We find their values by matching
the interior spacetime with the exterior geometry at the spherical
junction. Thus, we choose the Schwarzschild spacetime representing
solution to the field equations in vacuum as
\begin{equation}\label{15}
ds^{2}_{+}=-\jmath
dt^{2}+\frac{1}{\jmath}dr^{2}+r^{2}\big(d\theta^{2}+\sin^{2}\theta
d\phi^{2}\big),
\end{equation}
where $\jmath=1-\frac{2M}{r}$ and $M$ being the total exterior mass.
In traditional GR, the exterior solution indeed simplifies to the
Schwarzschild metric when the source terms vanish. However, in
modified gravity theories, especially those incorporating additional
curvature and energy-momentum tensor terms, the correspondence to
the Schwarzschild solution might not be as straightforward, and
additional geometric terms may persist. Several researchers
discussed that even when matter and pressure tend towards zero,
certain geometrical terms may persist due to the intricate coupling
introduced by the modified gravity frameworks \cite{30,32}. Some
different but interesting works are \cite{72}-\cite{75}. The smooth
matching at the surface boundary $(\Sigma:~r=\mathrm{R})$ yields
\begin{eqnarray}\label{16}
A&=&\frac{\mathrm{R}-2M}{\mathrm{R}(1+B\mathrm{R}^{2})^{4}},\\\label{16a}
B&=&\frac{6\mathrm{R}^{3}-7M\mathrm{R}^{2}-2\sqrt{9\mathrm{R}^{6}-14M\mathrm{R}^{5}}}{7M\mathrm{R}^{4}-4\mathrm{R}^{5}}.
\end{eqnarray}
We consider estimated masses and radii of six different star
candidates in Table \textbf{1} that would be helpful to calculate
the above two constants and thus to plot the physical
characteristics of the resulting solution. We also provide a
constant doublet $(A,B)$ corresponding to each candidate in Table
\textbf{1}. The components \eqref{11} and \eqref{11a} are plotted in
Figure \textbf{1}, showing non-singular and increasing profile
everywhere in the considered range of celestial objects.
\begin{figure}\center
\epsfig{file=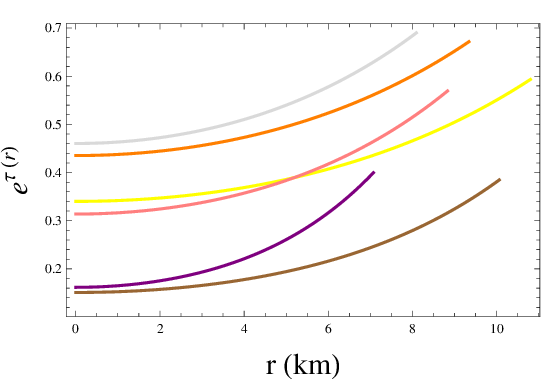,width=.5\linewidth}\epsfig{file=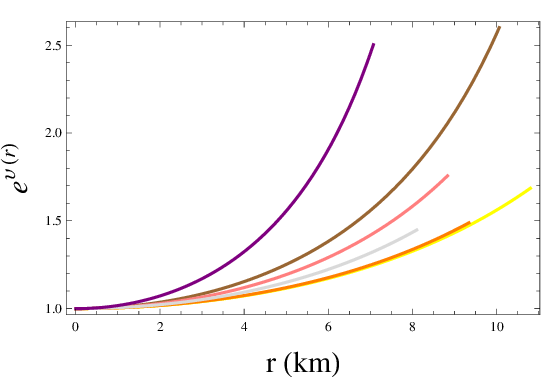,width=.5\linewidth}\caption{Metric
potentials versus $r$ corresponding to SMC X-1 (orange), PSR
J0348-0432 (brown), Her X-1 (gray), LMC X-4 (cyan), Cen X-3 (yellow)
and SAX J 1808.4-3658 (purple).}
\end{figure}
Since the matter variables are appeared up to quadratic order in the
field equations, their explicit expressions cannot be obtained.
Therefore, we consider $\rho=\frac{3m}{4\pi r^{3}}$,
$P_{r}=\frac{\rho}{3}$ and $P_{t}=(\varsigma+1)P_{r}$ with $m=m(r)$
being the interior mass and $\varsigma>0$. It must be mentioned here
that the pressure anisotropy in the interior of a self-gravitating
model must be positive in order to produce enough outward pressure
to counterbalance the inward gravitational force. We observe that
the above considered form of the anisotropic function is aligned
with such scenarios. Inserting these assumptions in the field
equations, they become
\begin{align}\nonumber
\rho^{eff}&=\frac{1}{f_{\mathcal{R}}}\bigg[\frac{3m}{4\pi
r^{3}}+\frac{m^{2}}{12\pi^{2}r^{6}}(\varsigma^{2}+9\varsigma+18)f_{\mathrm{T}^{2}}+\frac{7-B^{2}r^{4}-10Br^{2}}
{7(1+Br^{2})^{2}}\\\label{12}
&\times\bigg\{f''_{\mathcal{R}}-\bigg(\frac{8Br(-3+Br^{2})}{(1+B
r)(-7+10Br^{2}+B^{2}r^{4})}-\frac{2}{r}\bigg)f'_{\mathcal{R}}\bigg\}
-\frac{f-\mathcal{R}f_{\mathcal{R}}}{2}\bigg],\\\nonumber
P^{eff}_{r}&=\frac{1}{f_{\mathcal{R}}}\bigg[\frac{m}{4\pi
r^{3}}+\frac{m^{2}}{12\pi^{2}r^{6}}(\varsigma-\varsigma^{2})f_{\mathrm{T}^{2}}-\frac{7-B^{2}r^{4}-10Br^{2}}
{7(1+Br^{2})^{2}}\\\label{13}&\times\bigg(\frac{2}{r}+\frac{4Br}{1+Br^{2}}\bigg)f'_{\mathcal{R}}
+\frac{1}{2}(f-\mathcal{R}f_{\mathcal{R}})\bigg],\\\nonumber
P^{eff}_{t}&=\frac{1}{f_{\mathcal{R}}}\bigg[(1+\varsigma)\frac{m}{4\pi
r^{3}}-\frac{m^{2}\varsigma}{24\pi^{2}r^{6}}f_{\mathrm{T}^{2}}+\frac{1}{2}(f-\mathcal{R}f_{\mathcal{R}})-\frac{7-B^{2}r^{4}
-10Br^{2}}{7(1+Br^{2})^{2}}\\\label{14}
&\times\bigg\{f''_{\mathcal{R}}+\bigg(\frac{4Br}{1+Br^{2}}-\frac{8Br(-3+Br^{2})}{(1+B
r)(-7+10Br^{2}+B^{2}r^{4})}+\frac{1}{r}\bigg)f'_{\mathcal{R}}\bigg\}\bigg].
\end{align}

\section{Different Models of Modified Gravity}

The inclusion of multivariate functions in the field equations
\eqref{12}-\eqref{14} makes this theory different from others, and
thus these equations are much complicated to be solved. In order to
tackle this situation, we focus on two particular models within the
framework of EMSG theory. Since such kind of matter-geometry coupled
theories can be discussed by taking either minimal or non-minimal
models, we specifically assume the former model in which geometry
and matter distribution are independent of each other. Such models
in this theory can be provided by
\begin{equation}\label{17}
f(\mathcal{R},\mathrm{T}^{2})=f_{1}(\mathcal{R})+f_{2}(\mathrm{T}^{2}).
\end{equation}
Several EMSG models have been discussed in the literature by taking
different forms of $f_{1}(\mathcal{R})$ and $f_{2}(\mathrm{T}^{2})$.
In the following, we shall discuss two different forms of
$f_{1}(\mathcal{R})$ while fixing $f_{2}(\mathrm{T}^{2})=\beta
\mathrm{T}^{2}$ with $\beta$ being an arbitrary constant.

\subsection*{Model I}

Here, we adopt a particular form of $f_{1}(\mathcal{R})$ proposed by
Starobinsky \cite{54}. Thus, the model \eqref{17} takas the form
\begin{equation}\label{18}
f(\mathcal{R},\mathrm{T}^{2})=\mathcal{R}+\alpha
\mathcal{R}^{2}+\beta \mathrm{T}^{2},
\end{equation}
where $\alpha$ is a non-negative model parameter. Notice that
$\beta=0$ reduces the results of this theory to those in
$f(\mathcal{R})$ framework. Moreover, $\alpha=0=\beta$ leads them to
GR. Equations \eqref{12}-\eqref{14} become under this model as
\begin{align}\nonumber
\rho^{eff}&=\frac{1}{1+2\alpha\mathcal{R}}\bigg[\frac{3m}{4\pi
r^{3}}+\frac{\beta m^{2}}{48
\pi^{2}r^{6}}\big(54+33\varsigma+4\varsigma^{2}\big)+\frac{7-B^{2}r^{4}-10Br^{2}}{7(1+Br^{2})^{2}}
\\\label{19}
&\times\bigg\{2\alpha\mathcal{R}''-2\alpha\mathcal{R}'\bigg(\frac{8Br(-3+Br^{2})}{(1+B
r)(-7+10Br^{2}+B^{2}r^{4})}-\frac{2}{r}\bigg)\bigg\}+\frac{1}{2}\alpha\mathcal{R}^{2}\bigg],
\\\nonumber
P^{eff}_{r}&=\frac{1}{1+2\alpha \mathcal{R}}\bigg[\frac{m}{4\pi
r^{3}}+\frac{\beta m^{2}}{48
\pi^{2}r^{6}}\big(18+7\varsigma-4\varsigma^{2}\big)-\frac{7-B^{2}r^{4}-10Br^{2}}{7(1+Br^{2})^{2}}
\\\label{20}
&\times\bigg(\frac{4Br}{1+Br^{2}}+\frac{2}{r}\bigg)2\alpha\mathcal{R}'
-\frac{1}{2}\alpha\mathcal{R}^{2}\bigg],
\\\nonumber
P^{eff}_{t}&=\frac{1}{1+2\alpha
\mathcal{R}}\bigg[\big(1+\varsigma\big)\frac{m}{4\pi
r^{3}}+\frac{\beta
m^{2}}{16\pi^{2}r^{6}}\big(6+\frac{1}{3}\varsigma\big)-\frac{7-B^{2}r^{4}-10Br^{2}}{7(1+Br^{2})^{2}}
\\\nonumber
&\times\bigg\{2\alpha\mathcal{R}'' +2\alpha
\mathcal{R}'\bigg(\frac{4Br}{1+Br^{2}}-\frac{8Br(-3+Br^{2})}{(1+B
r)(-7+10Br^{2}+B^{2}r^{4})}+\frac{1}{r}\bigg)\bigg\}
\\\label{21}
&-\frac{1}{2} \alpha\mathcal{R}^{2}\bigg].
\end{align}

\subsection*{Model II}

In this subsection, we choose another form of $f_{1}(\mathcal{R})$
as
$\mathcal{R}-\omega\upsilon\tanh\left(\frac{\mathcal{R}}{\omega}\right)$
with a positive constant $\omega$ and non-negative $\upsilon$
\cite{55}. The model \eqref{17} thus turns into
\begin{equation}\label{22}
f(\mathcal{R},\mathrm{T}^{2})=\mathcal{R}-\upsilon\omega\tanh(\varphi)+\beta\mathrm{T}^{2},
\end{equation}
where $\varphi=\frac{\mathcal{R}}{\omega}$. Combining the above
model with Eqs.\eqref{12}-\eqref{14}, we obtain
\begin{align}\nonumber
\rho^{eff}&=\frac{1}{1-\upsilon\sec
h^{2}\varphi}\bigg[\frac{\upsilon\omega}{2}\tanh^{2}\varphi+\frac{3m}{4\pi
r^{3}}+\frac{\beta m^{2}}{48
\pi^{2}r^{6}}\big(54+33\varsigma+4\varsigma^{2}\big)-\frac{\upsilon
\mathcal{R}}{2}
\\\nonumber
&\times\sec
h^{2}\varphi+\frac{7-B^{2}r^{4}-10Br^{2}}{7(1+Br^{2})^{2}}
\bigg\{\frac{\upsilon \sec
h^{4}\varphi}{\omega^{2}}(2-4\tanh^{2}\varphi)\mathcal{R}'^{2}+\frac{\sec
h^{4}\varphi}{\omega}
\\\label{23}
&\times\upsilon\bigg(\mathcal{R}''-\mathcal{R}'\bigg(\frac{8Br(-3+Br^{2})}{(1+B
r)(10Br^{2}-7+B^{2}r^{4})}-\frac{2}{r}\bigg)\bigg)\bigg\}\bigg],
\\\nonumber
P^{eff}_{r}&=\frac{1}{1-\upsilon\sec h^{2}\varphi}\bigg[\frac{\beta
m^{2}}{48\pi^{2}r^{6}}\big(18+7\varsigma-4\varsigma^{2}\big)+\frac{m}{4\pi
r^{3}}+\upsilon\sec h^{2}\varphi\bigg\{\frac{\mathcal{R}}{2}
\\\label{24}
&-\frac{7-B^{2}r^{4}-10Br^{2}}{7(1+Br^{2})^{2}}
\bigg(\frac{4Br}{1+Br^{2}}+\frac{2}{r}\bigg)\frac{2\tanh\varphi
\mathcal{R}'}{\omega}\bigg\}-\frac{\upsilon\omega}{2}\tanh\varphi\bigg],
\\\nonumber
P^{eff}_{t}&=\frac{1}{1-\upsilon\sec h^{2}\varphi}\bigg[\frac{\beta
m^{2}}{16\pi^{2}r^{6}}\big(6+\frac{1}{3}\varsigma\big)+\big(1+\varsigma
\big)\frac{m}{4\pi r^{3}}-\frac{7-B^{2}r^{4}
-10Br^{2}}{7(1+Br^{2})^{2}}
\\\nonumber
&\times\bigg\{(2-4\tanh^{2}
\varphi)\mathcal{R}'^{2}+\omega\sinh2\varphi\bigg(\mathcal{R}''
+\mathcal{R}'\bigg(\frac{4Br}{1+Br^{2}}-\frac{8Br}{(1+Br)}
\\\label{25}
&\times\frac{(Br^{2}-3)}{(10Br^{2}-7+B^{2}r^{4})}+\frac{1}{r}\bigg)\bigg)\bigg\}
\frac{\upsilon\sec
h^{4}\varphi}{\omega^{2}}-\frac{\upsilon\omega}{2}\tanh \varphi
+\frac{\upsilon\mathcal{R}}{2}\sec h^{2}\varphi\bigg].
\end{align}

\section{Physical Features of Stellar Objects}

In this section, we explore some physical features of the considered
anisotropic star candidates through graphical analysis. We
investigate varying profiles of several parameters including the
effective matter determinants, EoS parameters, viability conditions,
surface redshift and compactness within the interior of stars for
both models I and II by choosing $\alpha=0.1$, $\beta=0.2$,
$\varsigma=1.5$, $\upsilon=0.01$ and $\omega=0.2$. Furthermore, the
stability shall be evaluated through the sound speed and cracking
approaches in the next section.

Here, we opt to fix the model parameters as a methodological choice
to maintain a focused exploration of the anisotropic star solutions
in modified gravity. The decision to set these parameters as
constants is primarily motivated by the desire to isolate the
effects of anisotropy and modified gravity on the star's structure.
While we acknowledge that alternative approaches, such as
constraining the parameter values through energy conditions or other
information, are valid and have been employed in various studies,
our decision to fix the parameters is in line with certain
methodological precedents within the literature. We have observed
that similar analysis in the field often adopt this approach to
simplify the investigation and draw clearer conclusions about the
specific aspects under consideration.

\subsection{Effective Matter Determinants}

Fluid parameters are fundamental quantities used to describe and
characterize the behavior of matter in celestial objects such as
stars and galaxies. They must be maximum in the core and exhibit a
gradually decreasing profile as we approach to the surface of
neutron stars. This behavior is a direct consequence of the dense
nature of these stellar objects. The graphical representation in
Figure \textbf{2} illustrates that the effective energy density and
pressure components possess a well-agreed trend corresponding to
both modified models. Furthermore, their derivatives with respect to
the radial coordinate are shown to be consistent with the regular
conditions. Their graphs are provided in Figure \textbf{3}.
\begin{figure}\center
\epsfig{file=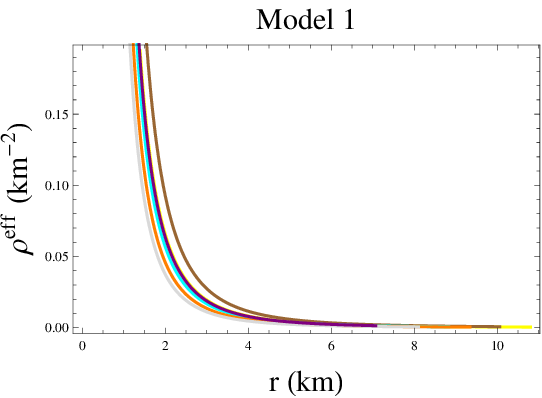,width=.5\linewidth}\epsfig{file=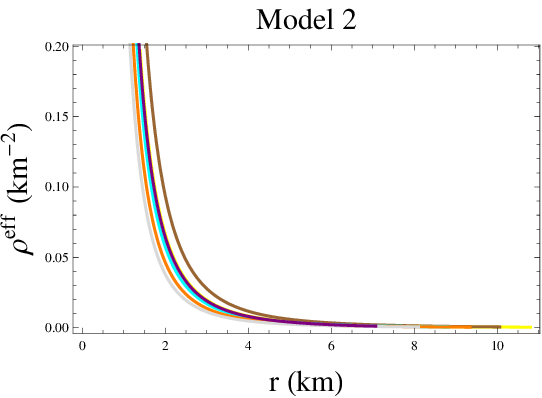,width=.5\linewidth}
\epsfig{file=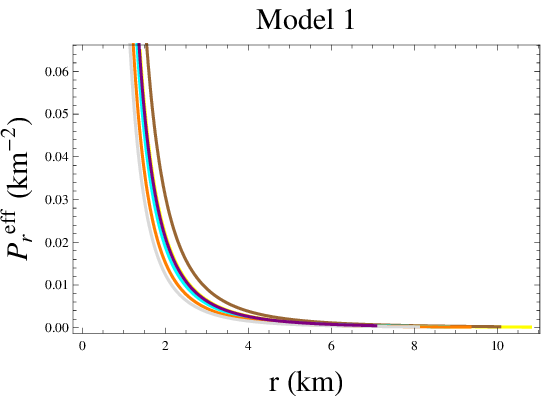,width=.5\linewidth}\epsfig{file=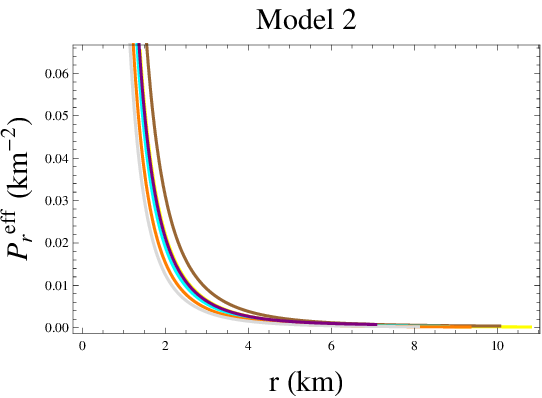,width=.5\linewidth}
\epsfig{file=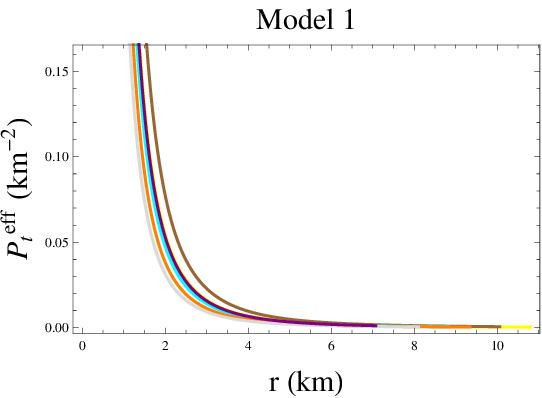,width=.5\linewidth}\epsfig{file=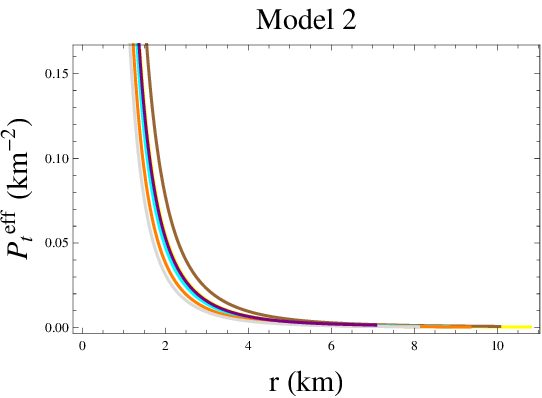,width=.5\linewidth}
\caption{Effective matter determinants versus $r$ corresponding to
SMC X-1 (orange), PSR J0348-0432 (brown), Her X-1 (gray), LMC X-4
(cyan), Cen X-3 (yellow) and SAX J 1808.4-3658 (purple).}
\end{figure}
\begin{figure}\center
\epsfig{file=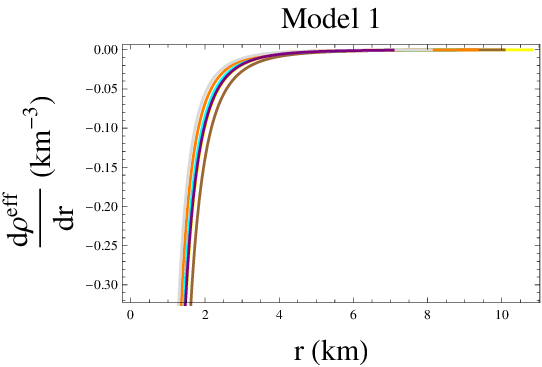,width=.5\linewidth}\epsfig{file=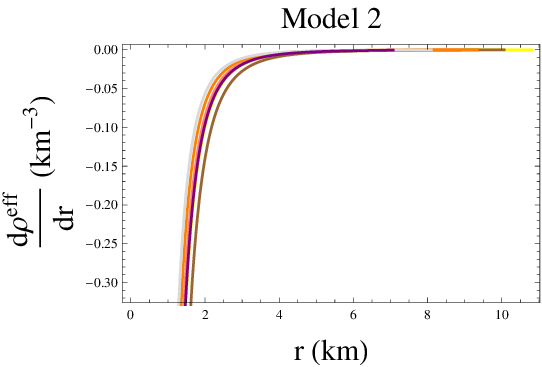,width=.5\linewidth}
\epsfig{file=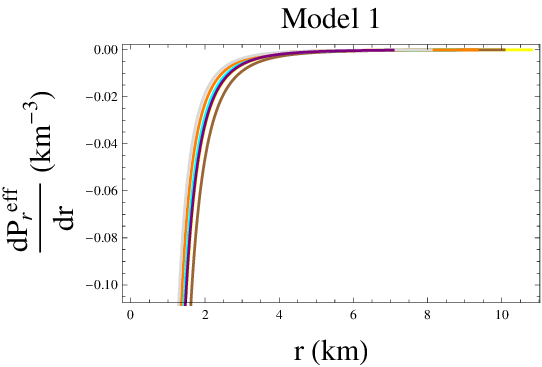,width=.5\linewidth}\epsfig{file=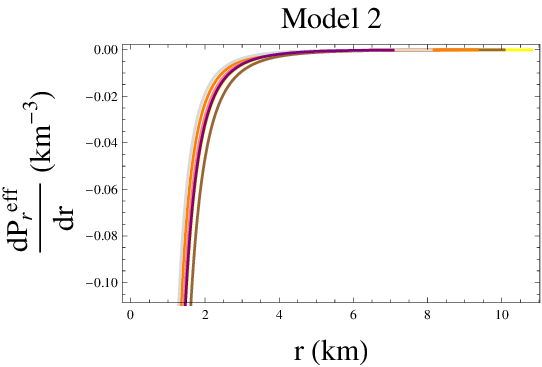,width=.5\linewidth}
\epsfig{file=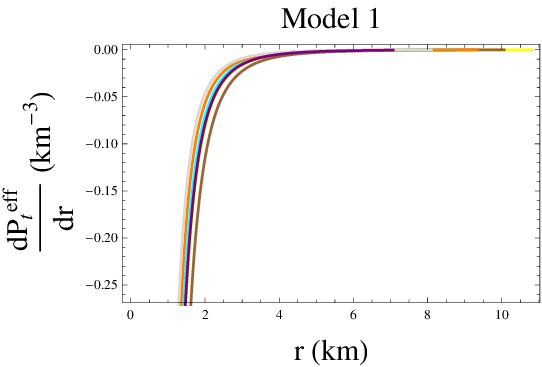,width=.5\linewidth}\epsfig{file=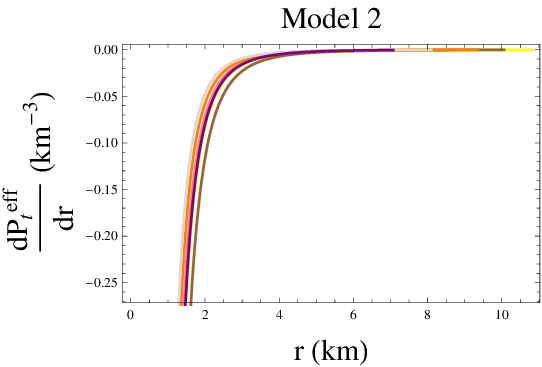,width=.5\linewidth}
\caption{Gradients of effective matter determinants versus $r$
corresponding to SMC X-1 (orange), PSR J0348-0432 (brown), Her X-1
(gray), LMC X-4 (cyan), Cen X-3 (yellow) and SAX J 1808.4-3658
(purple).}
\end{figure}

\subsection{Anisotropy}

Anisotropy means pressure within the system varies with respect to
the direction along which it is measured. Determining whether the
anisotropic pressure is positive or negative is a crucial factor in
discussing the stability of stellar objects.
\begin{itemize}
\item When the anisotropy is positive, it indicates that pressure
is directed outward from the center of the system, counterbalancing
the inward-directed gravitational force. This phenomenon can be
observed in various physical systems, such as stars or in other
bodies where the pressure acts to expand or push outward in specific
directions.
\item When the anisotropy is negative, it produces the inward-directed
pressure that would lead to unstable objects. Negative anisotropy is
observed in situations where the pressure acts to contract or pull
inward along certain axes within the system.
\end{itemize}
In Figure \textbf{4}, the anisotropy
($\Delta^{eff}=P^{eff}_{t}-P^{eff}_{r}$) is positively oriented,
indicating a situation where the pressure within the system exerts
an outward force.
\begin{figure}\center
\epsfig{file=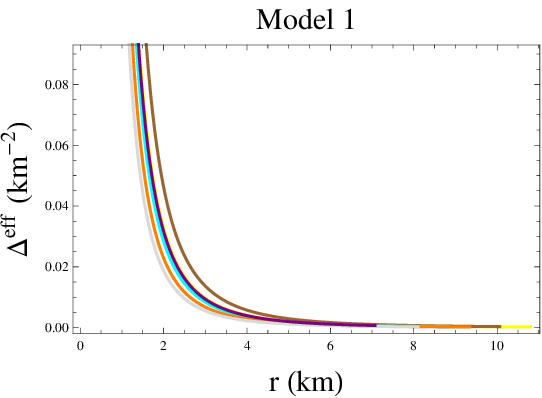,width=.5\linewidth}\epsfig{file=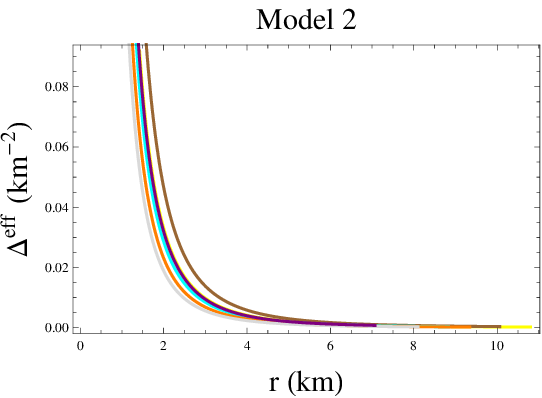,width=.5\linewidth}
\caption{Anisotropic factor versus $r$ corresponding to SMC X-1
(orange), PSR J0348-0432 (brown), Her X-1 (gray), LMC X-4 (cyan),
Cen X-3 (yellow) and SAX J 1808.4-3658 (purple).}
\end{figure}

\subsection{Energy Conditions}

The energy conditions are of great importance because they help in
determining the nature of the matter within the self-gravitating
interior. For instance, their fulfillment confirms that the interior
must hold an ordinary matter, otherwise, it exhibits some exotic
properties. Additionally, they are used to check the viability of
the solutions to the field equations. In other words, these
conditions allow researchers to study whether the developed
theoretical solutions in GR or other modified theories are
physically realistic or not. The energy conditions are categorized
as null $NECs$, strong $SECs$, dominant $DECs$ and weak $WECs$
defined as
\begin{itemize}
\item $NECs$
\\\\
$0\leq P^{eff}_{r}+\rho^{eff}, \quad 0\leq P^{eff}_{t}+\rho^{eff}$,
\item $SECs$
\\\\
$0\leq P^{eff}_{r}+\rho^{eff}, \quad 0\leq P^{eff}_{t}+\rho^{eff},
\quad 0\leq 2P^{eff}_{t}+P^{eff}_{r}+\rho^{eff}$,
\item $DECs$
\\\\
$0\leq \rho^{eff}\pm P^{eff}_{r}, \quad 0\leq \rho^{eff}\pm
P^{eff}_{t}$,
\item $WECs$
\\\\
$0\leq\rho^{eff}, \quad 0\leq \rho^{eff}+P^{eff}_{t}, \quad 0 \leq
\rho^{eff}+P^{eff}_{r}$.
\end{itemize}
Figures \textbf{5} and {6} provide clear evidence that the
considered stars have the characteristics of an ordinary matter, as
all the energy constraints are fulfilled.
\begin{figure}\center
\epsfig{file=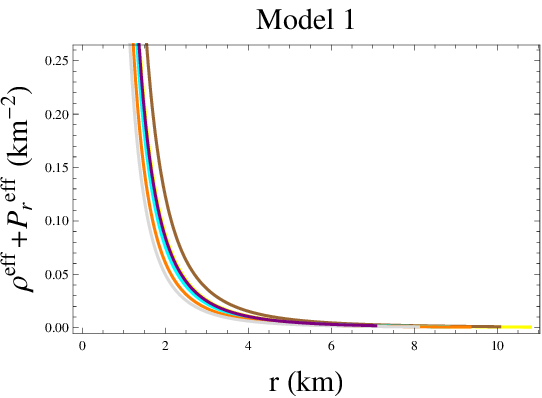,width=.5\linewidth}\epsfig{file=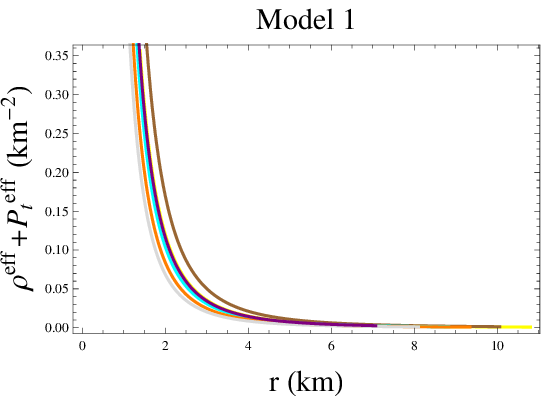,width=.5\linewidth}
\epsfig{file=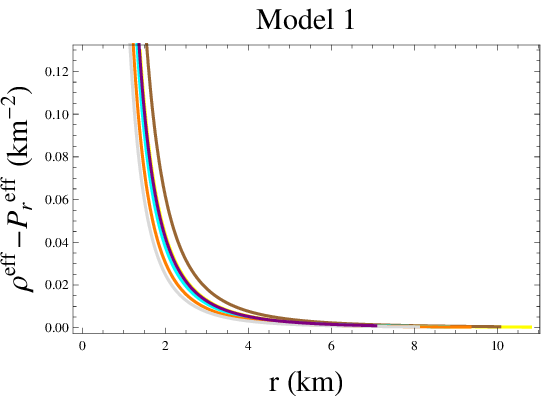,width=.5\linewidth}\epsfig{file=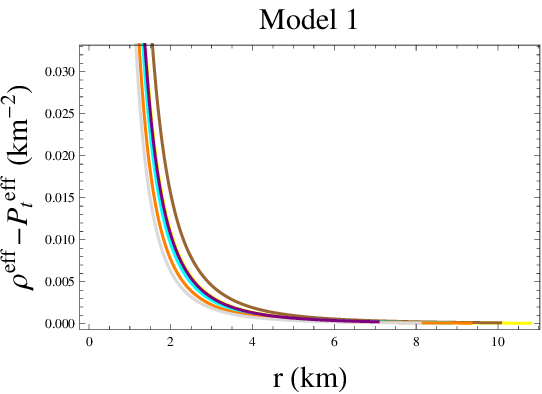,width=.5\linewidth}
\epsfig{file=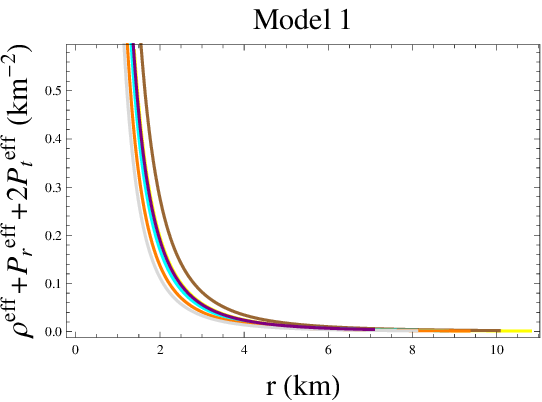,width=.5\linewidth} \caption{Energy bounds
versus $r$ corresponding to SMC X-1 (orange), PSR J0348-0432
(brown), Her X-1 (gray), LMC X-4 (cyan), Cen X-3 (yellow) and SAX J
1808.4-3658 (purple).}
\end{figure}
\begin{figure}\center
\epsfig{file=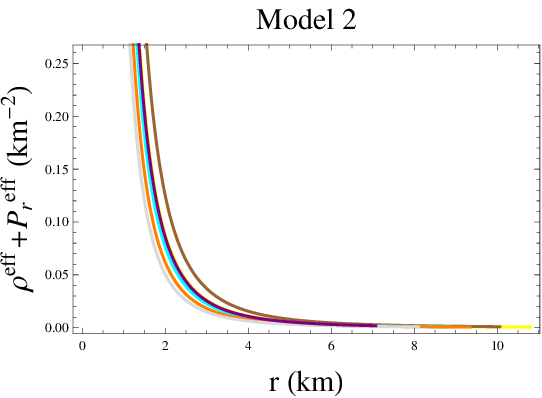,width=.5\linewidth}\epsfig{file=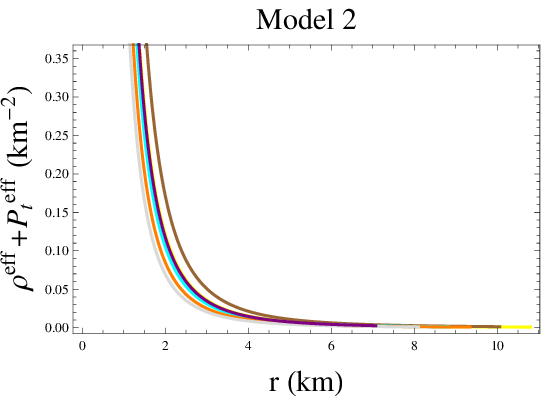,width=.5\linewidth}
\epsfig{file=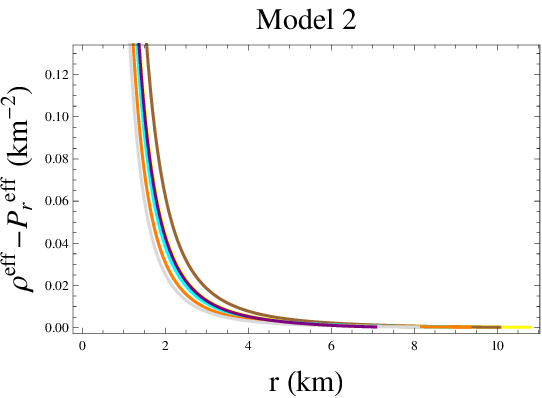,width=.5\linewidth}\epsfig{file=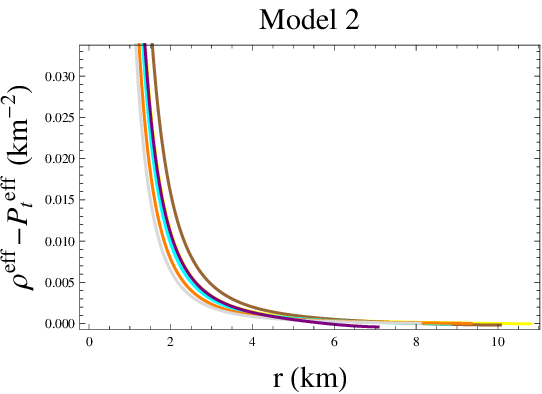,width=.5\linewidth}
\epsfig{file=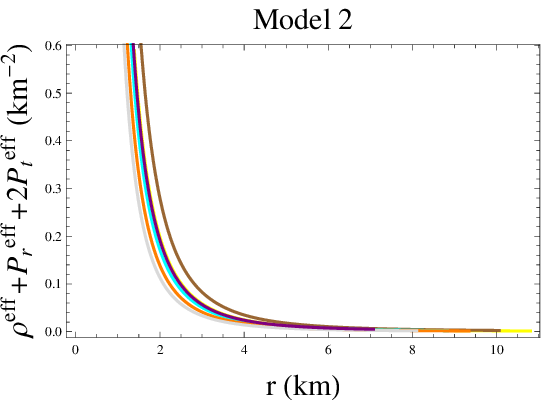,width=.5\linewidth} \caption{Energy bounds
versus $r$ corresponding to SMC X-1 (orange), PSR J0348-0432
(brown), Her X-1 (gray), LMC X-4 (cyan), Cen X-3 (yellow) and SAX J
1808.4-3658 (purple).}
\end{figure}

\subsection{Equation of State Parameters}

Here, we analyze some parameters to understand the relationship
between different fluid parameters within a system. An important
criteria for a physically feasible stellar model is that its EoS
parameters fall within the range $[0,1]$. Figure \textbf{7} shows
that the radial $\omega^{eff}_{r}$ and tangential $\omega^{eff}_{t}$
parameters defined below meet the required condition
$$\omega^{eff}_{r}=\frac{P^{eff}_{r}}{\rho^{eff}}, \quad
\omega^{eff}_{t}=\frac{P^{eff}_{t}}{\rho^{eff}}.$$ Saes and Mendes
\cite{45} described some interesting and useful properties of the
stiffness of nuclear matter through a particular EoS and discussed
its consequences in contempt of current as well as future
observations regarding neutron stars.
\begin{figure}[h!]\center
\epsfig{file=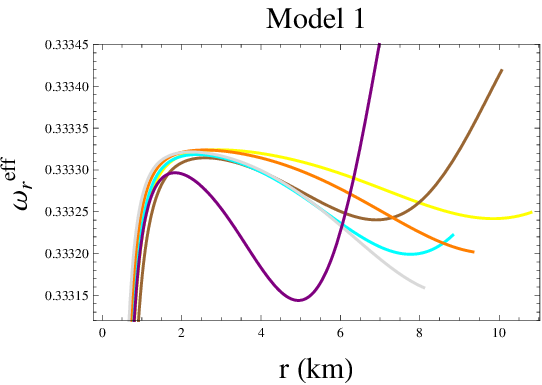,width=.5\linewidth}\epsfig{file=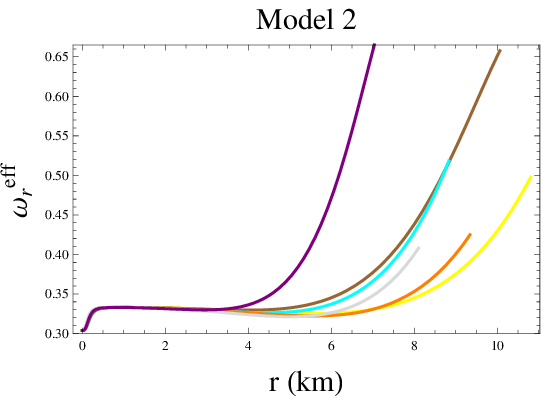,width=.5\linewidth}
\epsfig{file=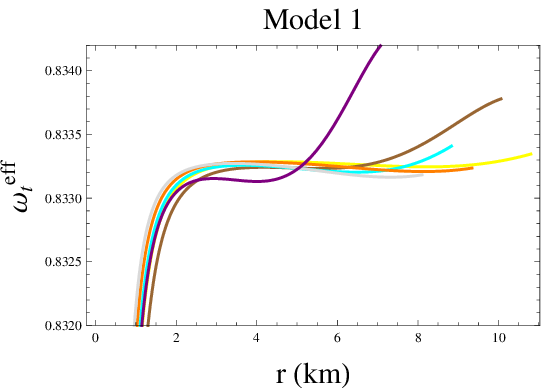,width=.5\linewidth}\epsfig{file=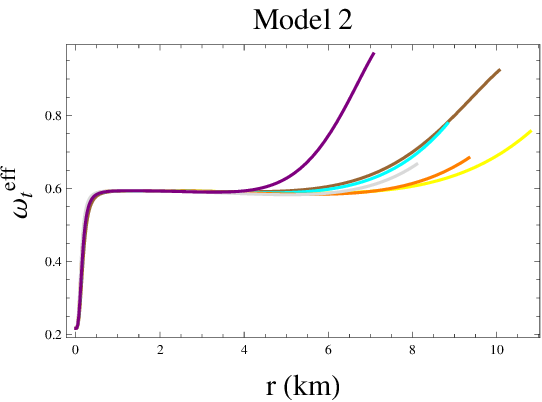,width=.5\linewidth}
\caption{EoS parameters versus $r$ corresponding to SMC X-1
(orange), PSR J0348-0432 (brown), Her X-1 (gray), LMC X-4 (cyan),
Cen X-3 (yellow) and SAX J 1808.4-3658 (purple).}
\end{figure}

\subsection{Mass, Compactness and Redshift}

Mass represents the total amount of matter contained within a given
volume or object. In stars, the mass is a fundamental parameter that
influences gravitational forces, determining the overall structure
and dynamics of that object. Mathematically, it is given as
\begin{equation}
m(r)=\frac{1}{2}\int_{0}^{\mathrm{R}}r^2\rho^{eff}dr.
\end{equation}
Figure \textbf{8} illustrates a positive correlation between mass
and radius, indicating that as the radius of the star increases, its
mass also increases. Compactness is a dimensionless parameter often
used to describe how tightly matter is packed within a celestial
object. It is defined as the ratio of mass to the object size
(typically expressed as its radius). This factor plays a role in
determining whether a celestial object will form a black hole after
collapsing or not. It can be expressed as
\begin{equation}
u(r)=\frac{m(r)}{r}.
\end{equation}
There is a specific limit for this factor proposed by Buchdhal
\cite{60} while discussing a physically relevant model. According to
him, this ratio should be less than $\frac{4}{9}$ everywhere. If
this ratio exceeds the suggested limit, then the structure may find
to be too dense or concentrated within its given radius, which could
lead to potential instability or collapse.

In cosmology, redshift is a measure of how much the light emitted by
distant objects has been stretched due to the expansion of the
universe. It provides crucial information about the relative motion
of celestial objects and is helpful in studying the large-scale
structures and expansion of the universe. We define it as follows
\begin{equation}
Z_{s}=\frac{1}{\sqrt{1-2u(r)}}-1.
\end{equation}
For a physically viable model, the redshift must be $Z_{s}\leq5.2$
\cite{61}. Figure \textbf{8} shows plots of these two factors which
are consistent with their respective findings.
\begin{figure}[h!]\center
\epsfig{file=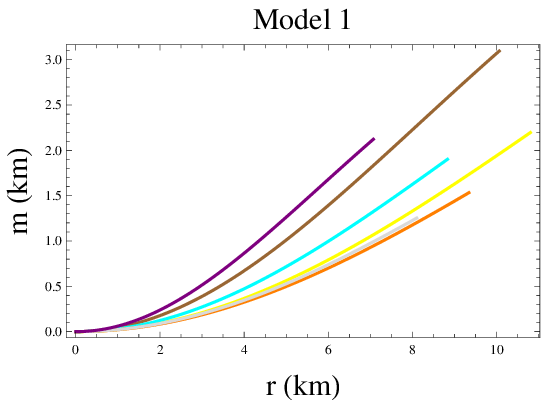,width=.5\linewidth}\epsfig{file=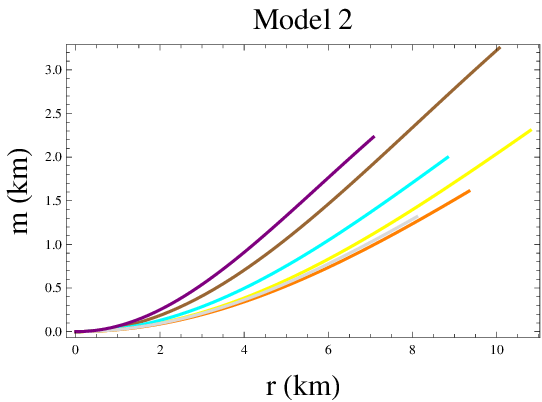,width=.5\linewidth}
\epsfig{file=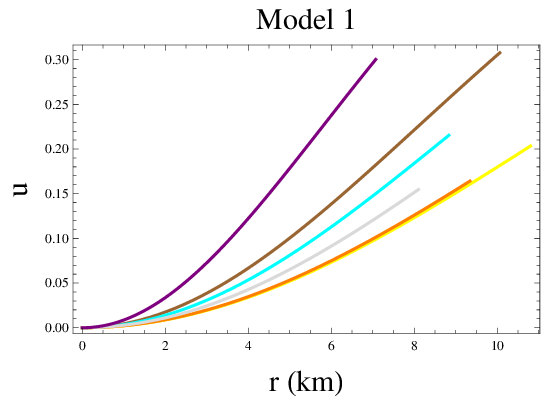,width=.5\linewidth}\epsfig{file=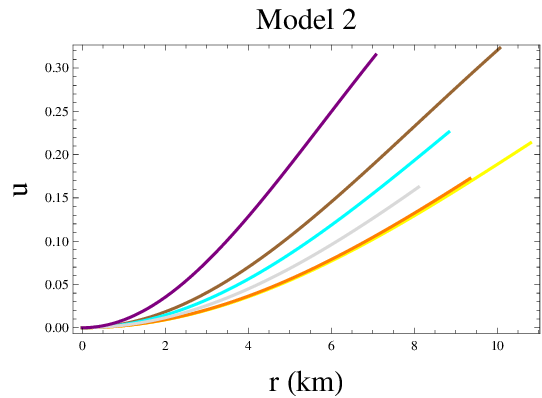,width=.5\linewidth}
\epsfig{file=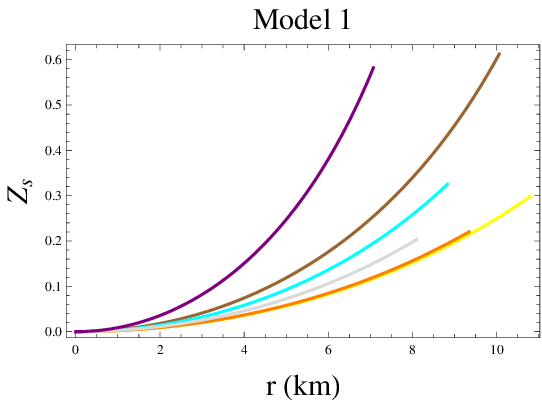,width=.5\linewidth}\epsfig{file=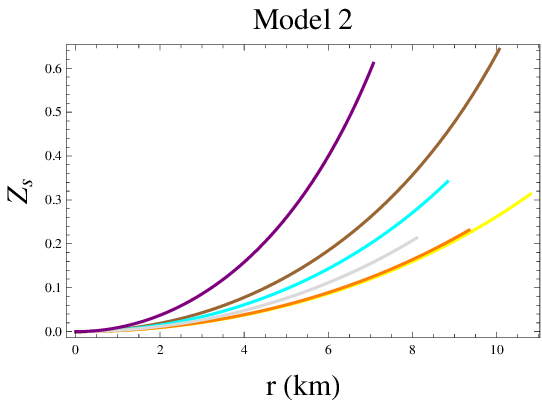,width=.5\linewidth}
\caption{Mass, compactness and redshift function versus $r$
corresponding to SMC X-1 (orange), PSR J0348-0432 (brown), Her X-1
(gray), LMC X-4 (cyan), Cen X-3 (yellow) and SAX J 1808.4-3658
(purple).}
\end{figure}

\section{Stability Analysis}

Stability is a fundamental concept to ensure the existence of
celestial objects. It becomes appealing to analyze the celestial
bodies that manage to maintain their stability even when subjected
to the external disturbances. In the realm of astrophysics, the
investigation of a star's stability often involves two important
notions, i.e., the sound speed, the cracking criterion and the
adiabatic index.

\subsection{Sound Speed}

Sound speed refers to the speed at which pressure waves (sound
waves) propagate through a medium. In the context of stars, sound
speed is a key indicator of how effectively pressure can counteract
gravitational forces. A stable star should have a sound speed
sufficient to maintain its structural integrity, preventing it from
the gravitational collapse. It also helps researchers to understand
how disturbances propagate within the star and whether they lead to
instability. According to the causality condition, it is necessary
that both the radial $v^{2(eff)}_{r}$ and transverse
$v_{t}^{2(eff)}$ components
$$v^{2(eff)}_{r}=\frac{dP^{eff}_{r}}{d\rho^{eff}}, \quad v_{t}^{2(eff)}=\frac{dP^{eff}_{t}}{d\rho^{eff}},$$
lie within the range of $(0,1)$ to get a stable interior. The
fulfilment of this criterion is verified in Figure \textbf{9}.

\subsection{Herrera's Cracking Approach}

The cracking approach is a theoretical framework pioneered by
Herrera and his colleagues \cite{50} to examine the stability of
self-gravitating systems. The interaction of pressure gradients with
the force of gravity produces instabilities in the system, leading
to the occurrence of cracking phenomenon. According to this
technique, a self-gravitating system can be considered stable only
when the difference between the radial and transverse components of
the sound speed should be in between 0 and 1. The failure of this
condition implies instability within the system, potentially leading
to a catastrophic collapse. Figure \textbf{9} ensures the stability
of all our considered stars (lower two plots).
\begin{figure}[h!]\center
\epsfig{file=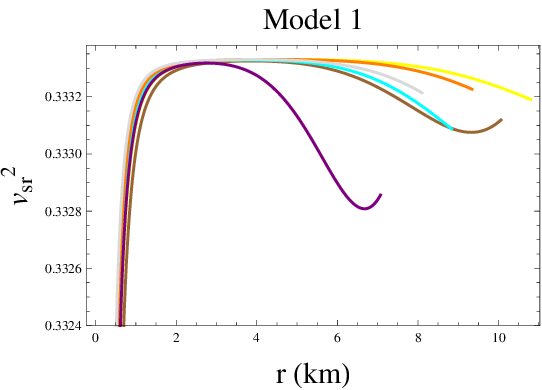,width=.5\linewidth}\epsfig{file=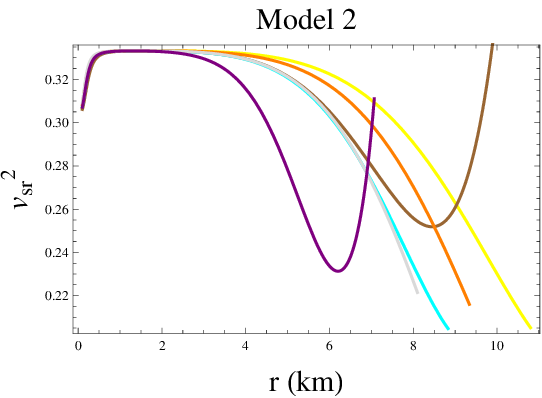,width=.5\linewidth}
\epsfig{file=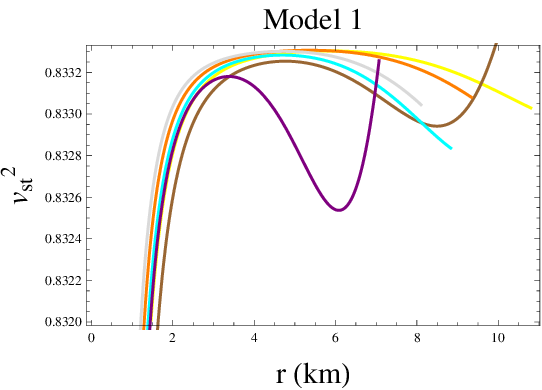,width=.5\linewidth}\epsfig{file=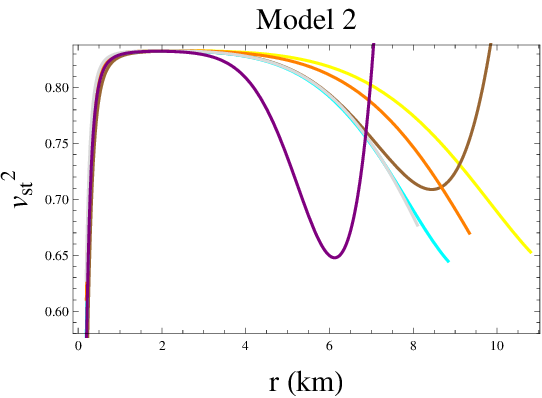,width=.5\linewidth}
\epsfig{file=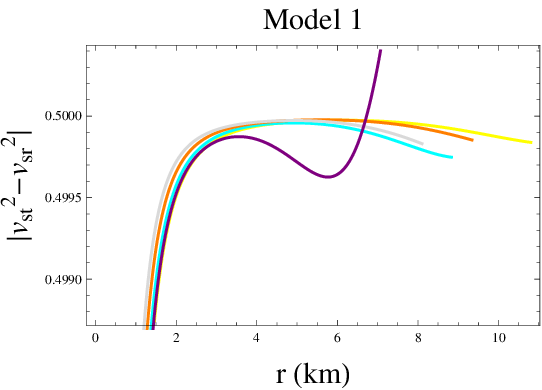,width=.5\linewidth}\epsfig{file=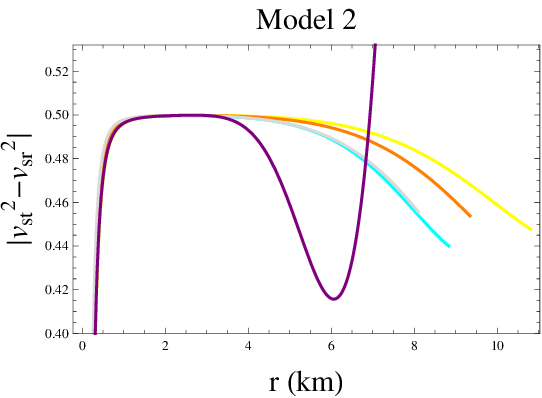,width=.5\linewidth}
\caption{Speed of sound and cracking versus $r$ corresponding to SMC
X-1 (orange), PSR J0348-0432 (brown), Her X-1 (gray), LMC X-4
(cyan), Cen X-3 (yellow) and SAX J 1808.4-3658 (purple).}
\end{figure}

\subsection{Adiabatic Index}

The adiabatic index is an important parameter used to study the
stability of self-gravitating objects. Stars are considered to be in
equilibrium state, where the inward force of gravity is
counterbalanced by the outward pressure generated by heat and
radiation inside the star. The stability of a star depends on the
balance between these two forces and the adiabatic index is a key
factor in determining the pressure force. According to Heintzmann
and Hillebrandt \cite{42e}, if the radial and transverse components,
respectively, defined by
$$\Gamma^{eff}_{r}=\frac{\rho^{eff}+P^{eff}_{r}}{P^{eff}_{r}}
\frac{dP^{eff}_{r}}{d\rho^{eff}},\quad
\Gamma^{eff}_{t}=\frac{\rho^{eff}+P^{eff}_{t}}{P^{eff}_{t}}
\frac{dP^{eff}_{t}}{d\rho^{eff}},$$ are greater than $\frac{4}{3}$,
then a perturbation compressing the star will cause an increase in
pressure that resists the compression leading to a stable star.
However, if the components of adiabatic index are less than the
defined limit, then the compression will cause a decrease in
pressure leading to further compression and instability. Figure
\textbf{10} shows our considered stars to be stable.
\begin{figure}
\epsfig{file=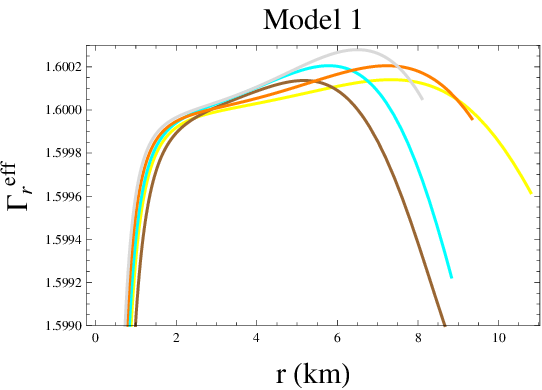,width=.5\linewidth}\epsfig{file=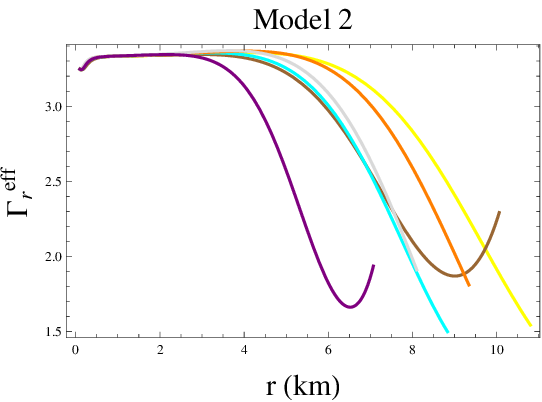,width=.5\linewidth}
\epsfig{file=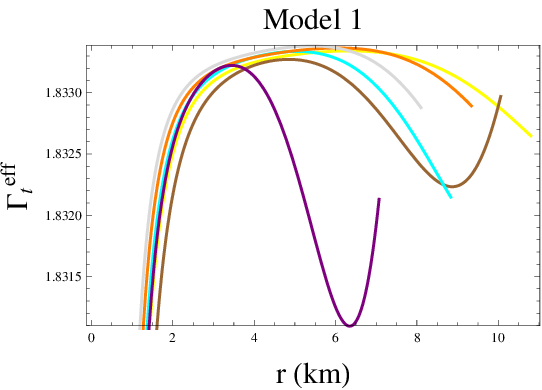,width=.5\linewidth}\epsfig{file=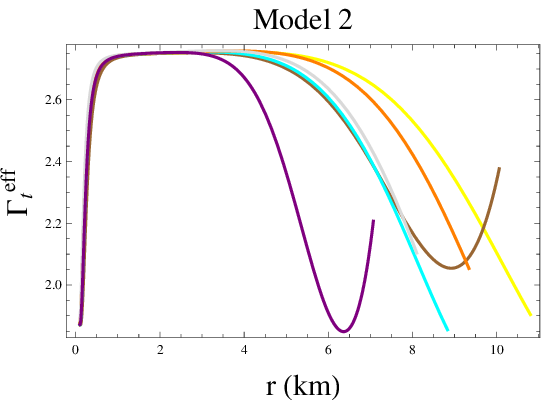,width=.5\linewidth}
\caption{Adiabatic index versus $r$ corresponding to SMC X-1
(orange), PSR J0348-0432 (brown), Her X-1 (gray), LMC X-4 (cyan),
Cen X-3 (yellow) and SAX J 1808.4-3658 (purple).}
\end{figure}

\section{Final Remarks}

This paper is devoted to the formulation of Durgapal-Fuloria
anisotropic solutions in the background of
$f(\mathcal{R},\mathrm{T}^{2})$ gravity. In order to do this, a
static spherical interior metric and the anisotropic energy-momentum
tensor have been considered. The corresponding modified field
equations have then been developed, containing extra degrees of
freedom. We have tackled with these equations by considering the
Durgapal-Fuloria metric possessing two constants $(A,B)$. These
unknowns have been calculated through matching conditions between
the interior and Schwarzschild exterior spacetimes. Further, we have
taken two different minimal models of this modified theory and
analyzed the physical behavior of our considered stars by examining
the resulting matter variables, anisotropy, energy conditions and
stability. The summary of our obtained results is presented as
follows.
\begin{itemize}
\item
Both the metric potentials are positive and exhibit increasing
profile. They show minimum value at the core of stars and
monotonically increasing behavior as we move towards the spherical
interface (Figure \textbf{1}).
\item
The effective matter variables reach their highest values at $r=0$
and gradually decrease outwards (Figure \textbf{2}). The derivatives
of these effective parameters indicate the presence of dense neutron
stars (Figure \textbf{3}).
\item
The positive anisotropy manifests the presence of a repulsive force
essential for the stability of neutron objects (Figure \textbf{4}).
\item
All energy constraints are satisfied providing a strong evidence of
the viability of our developed solutions. They also signify that the
interior of stellar objects primarily consists of an ordinary matter
(Figures \textbf{5} and \textbf{6}).
\item
Our considered models are consistent as the EoS parameters lie
within the range $[0,1]$ (Figure \textbf{7}).
\item
The behavior of mass function, compactness and redshift functions
have also found to be within their required limits (Figure
\textbf{8}).
\item Deriving the Tolman-Oppenheimer-Volkoff (TOV) equations for our
formulated models is a subject of great discussion when discussing
the neutron stars. However, we would like to mention the inherent
limitations of the modified gravitational theory we are working
within. Unfortunately, the specific form of the theory we are
investigating does not readily provide the graphical profiles of the
energy density and pressure components at the center of the massive
star unlike $f(\mathcal{R},\mathrm{T})$ \cite{52} and
$f(\mathcal{R},\mathrm{T},\mathcal{Q})$ (where
$\mathcal{Q}\equiv\mathcal{R}_{\xi\psi}\mathrm{T}^{\xi\psi}$)
\cite{53} theories. This limitation arises from the intricate nature
of the theory, and as a result, constructing plots of the TOV
equations may not be feasible within the scope of this study.
\item All the stability requirements have been satisfied
for chosen parametric values (Figures \textbf{9} and \textbf{10}).
\end{itemize}
It is important to mention here that the values of all physical
parameters in this theory increase as compared to GR \cite{57,59}
and other modified theories. We conclude that our considered neutron
stars are physically viable and stable for both
$f(\mathcal{R},\mathrm{T}^{2})$ models.\\\\
\noindent \textbf{Data Availability Statement:} This manuscript has
no associated data.

\end{document}